\documentstyle[aps,epsf]{revtex}
\begin{document}
\title{Clocking hadronization in relativistic heavy ion collisions with balance
functions}

\author{Steffen A. Bass, Pawel Danielewicz and Scott Pratt}

\address{Department of Physics and Astronomy and National Superconducting
Cyclotron Laboratory,\\ 
Michigan State University, East Lansing, MI 48824~~USA}

\date{\today} 

\maketitle

%\pacs{}

\begin{abstract}
A novel state of matter has been hypothesized to exist during the early stage
of relativistic heavy ion collisions, with normal hadrons not appearing until
several fm/c after the start of the reaction. To test this hypothesis,
correlations between charges and their associated anticharges are evaluated
with the use of balance functions. It is shown that late-stage hadronization is
characterized by tightly correlated charge-anticharge pairs when measured
as a function of relative rapidity.
\end{abstract}

\vspace*{10pt}

Relativistic heavy ion collisions produce mesoscopic regions of enormous energy
density, perhaps surpassing 3 GeV/fm$^3$ in Pb collisions at the CERN SPS
\cite{na49_energydensity,wa98_energydensity} with even higher energy densities
expected at RHIC.  At such energies hadronic degrees of freedom should be
replaced by quark-gluon degrees of freedom. Several experimental measurements
have been proposed as signals to the quark-gluon plasma\cite{reviews}. Among
these signals is an expected enhancement in strange-quark production which
should take place 5-10 fm/c into the collision when the local temperature has
dropped to near 160 MeV, but the system is still far from
freeze-out. Strangeness enhancement has indeed been observed in heavy ion
collisions \cite{strange_data}, but alternative hadronic explanations have also
been put forward assuming early-stage hadronization with medium
modifications, referred to as color ropes \cite{biro84,sorge92} or baryon
junctions \cite{vance}. In this paper the use of balance functions is proposed
as a means to determine whether quark production occurred at early times,
$\tau<1$ fm/c, or according to a late-stage hadronization scenario, see
e.g. \cite{raf82,koch86}.

Late-stage production of quarks could be attributed to three mechanisms:
formation of hadrons from gluons, conversion of the non-perturbative vacuum
energy into particles, or hadronization of a quark gas at constant
temperature. Hadronization of a quark gas should approximately conserve the net
number of particles due to the constraint of entropy conservation. Since
hadrons are formed of two or more quarks, creation of quark-antiquark pairs
should accompany hadronization. All three mechanisms for late-stage quark
production involve a change in the degrees of freedom. Therefore, any signal
that pinpoints the time where quarks first appear in a collision would provide
valuable insight into understanding whether a novel state of matter has been
formed and persisted for a substantial time.
The fact that the hadronic phase has a higher concentration of charges than
the QGP phase at the same entropy has been discussed in the context of 
charge fluctuations in \cite{fluc_papers}. 

The link between balance functions and the time at which quarks are created has
a simple physical explanation. Charge-anticharge pairs are created at the same
location in space-time, and are correlated in rapidity due to the strong
collective expansion inherent to a relativistic heavy ion collision. Pairs
created earlier can separate further in rapidity due to the higher initial
temperature and due to the diffusive interactions with other particles. The
balance function, which describes the momentum of the accompanying
antiparticle, quantifies this correlation.

The balance functions employed here are similar to observables used to
investigate hadronization in jets produced in $p\bar{p}$ or $e^+e^-$ collisions
\cite{ppdata,eedata}. The balance function describes the conditional
probability that a particle in the bin $p_1$ will be accompanied by a particle
of opposite charge in the bin $p_2$. We define the balance function,
\begin{equation}
\label{balancedef_eq}
B(p_2|p_1)\equiv\frac{1}{2}\left\{
\rho(b,p_2|a,p_1)-\rho(b,p_2|b,p_1)+\rho(a,p_2|b,p_1)-\rho(a,p_2|a,p_1)
\right\},
\end{equation}
where $\rho(b,p_2|a,p_1)$ is the conditional probability of observing a
particle of type $b$ in bin $p_2$ given the existence of a particle of type $a$
in bin $p_1$.  The label $a$ might refer to all negative kaons with $b$
referring to all positive kaons, or $a$ might refer to all hadrons with a
strange quark while $b$ refers to all hadrons with an antistrange quark. The
conditional probability $\rho(b,p_2|a,p_1)$ is generated by first counting the
number $N(b,p_2|a,p_1)$ of pairs that satisfy both criteria and dividing by the
number $N(a,p_1)$ of particles of type $a$ that satisfy the first criteria.
\begin{equation}
\label{rhodef_eq}
\rho(b,p_2,a,p_1)=\frac{N(b,p_2|a,p_1)}{N(a,p_1)}.
\end{equation}
Both sums run over all events, though pairs only involve particles from the
same event. 

An example of binning might be that $p_1$ refers to a measurement anywhere in
the detector, while $p_2$ refers to the relative rapidity $|y_b-y_a|$. Then the
balance function would be a function of $\Delta y$ only, and would represent
the probability that the balancing charges were separated by $\Delta y$ (in our
formalism we include a division by $\Delta y$ to express $B(\Delta y)$ as a
density).

The balance function is normalized to unity if $a/b$ refer to all particles
with a positive/negative globally conserved charge.
\begin{equation}
\sum_{p_2}B(p_2|p_1)=\frac{1}{2}\left\{M_b-(M_b-1)+M_a-(M_a-1)\right\}=1,
\end{equation}
where $M_a$ and $M_b$ are the average multiplicities of the $a$ and $b$
particles.  The normalization derives from the fact that for every extra
positive charge there exists one extra negative charge. If the acceptance
measures only a fraction of the charge, e.g. only kaons are measured and the
strangeness in hyperons is excluded, the balance function would sum to that
fraction. Balance functions can exploit any conserved charge: electric charge,
strangeness, baryon number or charm. The first two terms in
Eq. (\ref{balancedef_eq}) constitute the balance functions defined in several
analyses of $e^+e^-\rightarrow$ jets. By adding the last two terms the
normalization properties are retained even for the case where there is a
non-zero net charge, $M_a-M_b\ne 0$.

If many charges are present in the event, the balance function represents the
subtraction of two large numbers. However, large multiplicities also imply a
large number of pairs from which to calculate the balance function. Since the
number of uncorrelated pairs rises as the square of the multiplicity $M$, the
statistical error in calculating the numerators of the conditional
probabilities, which rises as the square root of the number of pairs, increases
linearly with $M$. Since the denominator also rises linearly with $M$, the
statistical error in the balance function is independent of multiplicity and is
principally determined by the number of events:
\begin{equation}
\sigma_B\propto\\\frac{1}{\sqrt{N_{\rm ev}}}.
\end{equation}
Thus, the baryon-antibaryon balance function which might involve a few dozen
antibaryons would require the same number of events as the electric-charge
balance function which might be constructed from a thousand particles.
Typically, $10^5$ events are required to determine a balance function with
statistical fluctuations at the level of $10^{-2}$.

Balance functions probe the dynamics of charge-anticharge pairs by quantifying
the degree to which the charges are correlated in momentum space given the
constraint of being created at the same space-time point in a system exhibiting
strong position-momentum correlations such as a relativistic collision where
source velocities might span several units of rapidity. In a globally
equilibrated system with no collective flow, there would exist no correlation
between the balancing charges, and the numerator in Eq. (\ref{rhodef_eq}) would
factorize. The width of the balance function would then correspond to the
extent of single-particle emission in momentum space.

To illustrate the way in which balance functions quantify the charge-anticharge
correlations, we consider a Bjorken boost-invariant parameterization
\cite{bjorken_boostinv} of a source expanding along the $z$ axis with a
collective velocity proportional to the position, $ v_{\rm coll}=z/t$.  All
intrinsic variables, such as density or temperature, depend only on the proper
time $\tau=(t^2-z^2)^{1/2}$. We first consider only direct production of
hadrons, as the possibility of hadrons coalescing from quarks is discussed
later in the paper. Particles and antiparticles of mass $m$ are generated in
pairs at the same point in space-time following a local thermal distribution,
and the relative rapidities are used to generate balance functions. The
characteristic width of the balance function is determined by the ratio of the
temperature to the mass. Non-relativistically, $\sigma_y=(2T/m)^{1/2}$, and
heavier particles are characterized by narrower balance functions. For
particles with masses much less than the temperature, the balance functions
become independent of the temperature.

\begin{figure}
\epsfxsize=0.45\textwidth 
\centerline{\epsfbox{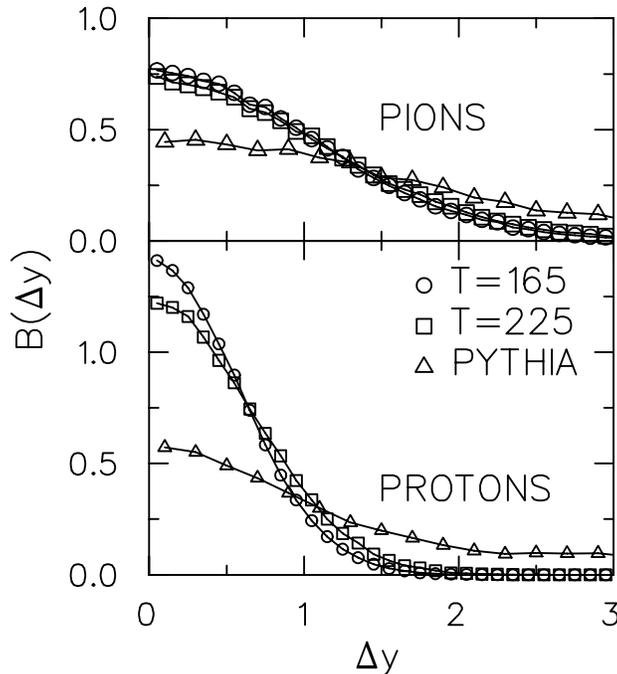}}
\caption{\label{masscompare_fig} Balance functions as predicted in a simple
Bjorken thermal model are shown for two temperatures, 225 MeV and 165 MeV.
Since heavier particles from cooler systems have smaller thermal velocities,
they are more strongly correlated in rapidity and result in narrower balance
functions.  Also shown are balance functions as predicted by PYTHIA where the
shape of the balance function is largely determined by string phenomenology.}
\end{figure}

Figure \ref{masscompare_fig} displays balance functions assuming a Bjorken
parameterization of an expanding pion gas and an expanding proton gas, for two
temperatures, 225 MeV and 165 MeV.  Clearly, the balance functions of the more
massive particles are sensitive to the temperature. This suggests that the
strangeness and baryon balance functions should provide more insight than the
electric-charge balance function which would be largely dominated by pions.

Balance functions in heavy ion collisions should be compared to those from $pp$
collisions at the same $\sqrt{s}$ where hadronization is nearly
instantaneous. Charged-pion balances measured in $e^+e^-$ collisions as a
function of the rapidity defined along the jet axis have been reasonably
explained by the string hadronization dynamics of the Lund model
\cite{lundmodel}, e.g. as implemented in PYTHIA \cite{pythia}. Thermally
generated balance functions are compared to predictions of PYTHIA for $pp$
collisions at $\sqrt{s}=200$ GeV in Fig. \ref{masscompare_fig}. The PYTHIA
balance functions tend to be broader than those that are thermally generated,
especially for the more massive protons and kaons. Assuming that experimental
balance functions in $pp$ collisions would be well described by similar string
dynamics, Fig. \ref{masscompare_fig} suggests that narrower balance functions
might indeed point to thermal production at a lower temperature and thus at
later times in the evolution of the heavy ion reaction.

Rescattering and annihilation should also affect balance functions.
Rescattering may be qualitatively understood by considering the diffusion
equation in Bjorken coordinates $\tau$ and $\eta\equiv\tanh^{-1}(z/t)$, where
$\eta$ plays the role of the position in the $z$ direction and also equals the
collective rapidity of the local matter. Rather than considering the diffusion
constant $D=v_t/(n\sigma)$ as a constant, it is more physical to incorporate
the fact that the density $n$ falls inversely with $\tau$ and to consider
$\beta\equiv v_t/(n\tau\sigma)$ as a constant where $v_t$ is the thermal
velocity and $\sigma$ is a characteristic cross section. The diffusion equation
then becomes
\begin{equation}
\frac{\partial}{\partial \tau}f(\tau,\eta)
=-\frac{\beta}{\tau}\frac{\partial^2}{\partial
\eta^2}f(\tau,\eta). 
\end{equation}
Here, $f$ is the probability of observing a particle at position $\eta$ at time
$\tau$. With the initial condition of $\eta=0$ at $\tau_0$, the solution to the
diffusion equation is a Gaussian with variance
$\sigma_\eta^2=2\beta\ln(\tau/\tau_0)$.  This illustrates that collisions
broaden the balance function by diffusing the charge in the effective spatial
coordinate $\eta$. However, in the limit of zero mean free path, the diffusion
constant tends to zero and the particles do not then diffuse.

The overall width of the balance function in relative rapidity is a combination
of the thermal rapidity spread $\sigma_{\rm therm}$ and the effect of
diffusion in $\eta$ of both particles:
\begin{equation}
\label{balsum_eq}
\sigma_y^2=\sigma_{\rm therm}^2+4\beta\ln(\tau/\tau_0).
\end{equation}
Due to cooling, the width $\sigma_{\rm therm}$ falls with time which provides a
competition between diffusion which stretches the balance function, and cooling
which narrows it. If the production occurs at early times, then
$\ln(\tau/\tau_0)$ is large and the effect of collisions is to significantly
broaden the balance function.

Some hadrons will contain coalesced quarks that were created at early
times. The thermal contribution to $\sigma_y$ described in
Eq. (\ref{balsum_eq}) should be unaffected by the past history of the
constituent quarks. However, the diffusive contribution might significantly
depend on the fact that the charge moved as a free quark rather than as a
hadron during it's early history. Balance functions constructed from hadrons
can thus provide meaningful information regarding the creation and mobility of
the constituent quarks.

\begin{figure}
\epsfxsize=0.45\textwidth 
\centerline{\epsfbox{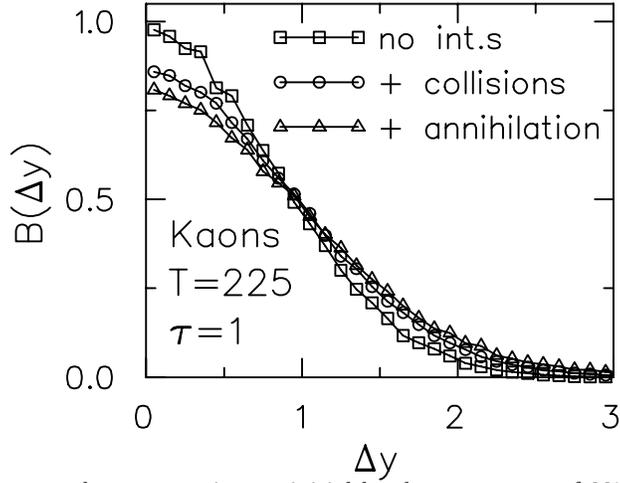}}
\caption{\label{collsanddeaths_fig} Kaon balance functions are shown assuming
an initial local temperature of 225 MeV and a production time of 1 fm/c. The
balance function is broadened by the inclusion of randomizing collisions and
annihilation.}
\end{figure}

To quantitatively illustrate the effect of rescattering, we model a pair of
particles produced at an initial proper time $\tau_0$ that collide $N_{\rm
coll}$ times before disassociating at a final time $\tau_f$. Each collision is
assumed to completely reorient the particle with the local collective
velocity. The collision times are chosen randomly such that the number of
collisions as a function of $\ln(\tau)$ is uniform. The temperature is chosen
to vary linearly with the proper time, cooling from 225 MeV at $\tau=1$ fm/c to
120 MeV at $\tau=15$ fm/c. Figure \ref{collsanddeaths_fig} shows the $K_+K_-$
balance function with $N_{\rm coll}=0$ and $N_{\rm coll}=10$ assuming kaons are
created at $\tau=1$ fm/c and cease to collide at $\tau_f=15$fm/c. In this case
collisions clearly broaden the balance function.

Annihilations should also broaden the balance function. Annihilation forms new
correlated pairs with the surviving partners of the annihilated particles,
which tend to be less correlated than the original pairs. Annihilation combined
with an equal amount of creation does not affect the balance function since the
relative rapidities of formed and annihilated pairs should be identical. Figure
\ref{collsanddeaths_fig} illustrates the effects of annihilation by considering
the same case described above, but with the additional assumption that half the
particles disappear due to annihilation. In hadronic models of heavy ion
collisions, the number of both antibaryons and strange particles tend to
decrease with time due to cooling, which should result in broadened balance
functions.

\begin{figure}
\epsfxsize=0.45\textwidth 
\centerline{\epsfbox{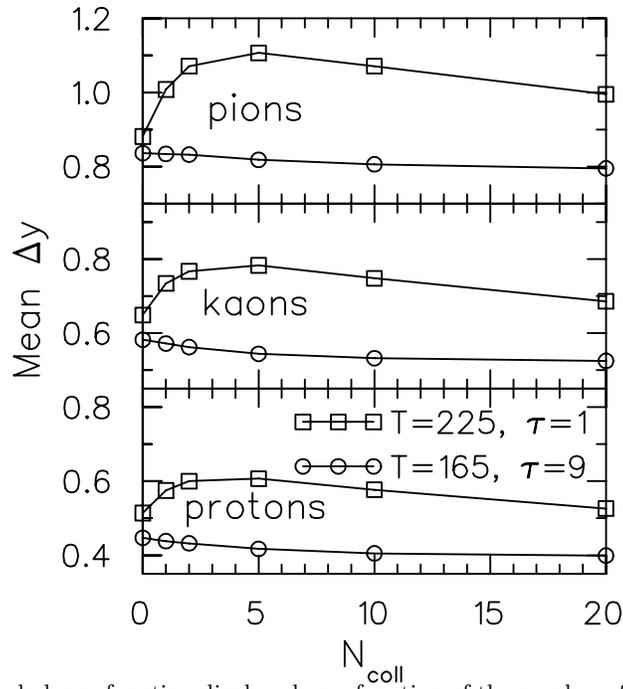}}
\caption{\label{collisions_fig} The mean width of the balance function
displayed as a function of the number of collisions, both for the case where
particles are created early ($\tau=1$ fm/c, $T=225$ MeV) and late ($\tau=9$
fm/c, $T=165$ MeV).}
\end{figure}

Figure \ref{collisions_fig} displays the effect of collisions on balance
functions for pions, kaons and protons, by considering the mean relative
rapidity as a function of the number of collisions. For production at early
times when the collective velocity gradient is high ($dv_{\rm
coll}/dz=1/\tau$), collisions broaden the balance function. However, for very
large numbers of collisions, the charge does not diffuse and the balance
functions are narrowed due to the cooling.  One would expect particles to
undergo 10-20 collisions if created at $\tau=1$ fm/c, although the effective
number of completely randomizing collisions might be closer to a half dozen. If
created at $\tau=9$ fm/c when the temperature is 165 MeV, the effective number
of completely randomizing collisions might be two or three. Figure
\ref{collisions_fig} suggests that the signal for late-stage quark production
is significantly magnified by rescattering. Due to collisions, even
charged-pion balance functions become strongly sensitive to the creation time.

The simple calculations presented here sidestep two issues: correlations from
decays such as $\phi\rightarrow K^+K^-$, and experimental acceptance
problems. Both problems can be addressed by modeling constrained by the
multitude of other observables measured in a heavy ion collision. Although some
open questions remain, it seems clear that the canonical picture of a heavy-ion
reaction,quark-gluon plasma formation followed by late-stage hadronization,
should have a clear signature in the balance functions. Compared to $pp$
collisions, one expects the peak in the balance function in nucleus-nucleus
collisions to be narrower near $\Delta y=0$ due to the contribution of
late-stage production of quark pairs, while the tails of balance function
should become broader reflecting the extra diffusion of charge in the early
stages of the collision. Finally, we remark that we have barely explored the
possibilities of balance functions. The rich nature of the binnings $(p_2|p_1)$
should provide a powerful means for resolving many of the issue regarding
creation and diffusion of quarks and hadrons in relativistic heavy ion
collisions.

\acknowledgements{We are grateful to T. Sj\"ostrand for providing valuable
references. This work was supported by the National Science Foundation, Grants
No. PHY-00-70818 and PHY-96-0527.}

\end{document}